# Ride-Hailing for Autonomous Vehicles: Hyperledger Fabric-Based Secure and Decentralize Blockchain Platform

Ryan Shivers*, Mohammad Ashiqur Rahman†, Md Jobair Hossain Faruk‡
Hossain Shahriar§, Alfredo Cuzzocrea¶, Victor Clincy∥
*Department of Computer Science, Tennessee Tech University, USA
†Department of Electrical and Computer Engineering, Florida International University, USA
‡Department of Software Engineering and Game Development, Kennesaw State University, USA
§Department of Information Technology, Kennesaw State University, USA
¶iDEA Lab, University of Calabria, Rende, Italy and LORIA, Nancy, France
∥Department of Computer Science, Kennesaw State University, USA
Email: rmshivers42@students.tntech.edu, marahman@fiu.edu, mhossa21@students.kennesaw.edu
hshahria@kennesaw.edu, alfredo.cuzzocrea@unical.it, vclincy@kennesaw.edu

*Abstract*—Ride-hailing and ride-sharing applications have recently gained popularity as a convenient alternative to traditional modes of travel. Current research into autonomous vehicles is accelerating rapidly and will soon become a critical component of a ride-hailing platform's architecture. Implementing an autonomous vehicle ride-hailing platform proves a difficult challenge due to the centralized nature of traditional ride-hailing architectures. In a traditional ride-hailing environment the drivers operate their own personal vehicles so it follows that a fleet of autonomous vehicles would be required for a centralized ride-hailing platform to succeed. Decentralization of the ride-hailing platform would remove a roadblock along the way to an autonomous vehicle ride-hailing platform by allowing owners of autonomous vehicles to add their vehicle to a community-driven fleet when not in use. Blockchain technology is an attractive choice for this decentralized architecture due to its immutability and fault tolerance. This thesis proposes a framework for developing a decentralized ride-hailing architecture that is verifiably secure. This framework is implemented on the Hyperledger Fabric blockchain platform. The evaluation of the implementation is done by applying known security models, utilizing a static analysis tool, and performing a performance analysis under heavy network load.

## I. INTRODUCTION

Ride-sharing services fill empty seats in cars with people who are traveling near the same destination as a driver. This concept of ride-sharing has evolved since its inception to a large-scale market as mass appeal has skyrocketed its profitability. The ride-sharing / ride-hailing marketplace has been rapidly expanding since the launch of companies such as Uber and Lyft that offer a platform for cooperation between riders and drivers through mobile applications. Goldman Sachs has predicted that the ride-sharing market revenue will be worth approximately 285 billion dollars by the year 2030 [1]. This prediction is not applicable to the market as it is now. However, self-driving car technology continues to advance at the rate predicted by Goldman Sachs analysts.

Autonomous Vehicles (AVs) collect information about the current state of the environment around them using sensors (*e.g.*, cameras, lasers, and electromagnetic field detectors) and feed the information into traditional Artificial Neural Networks to make decisions about how to operate the vehicle while on the road. The first system that operated in this manner was the Autonomous Land Vehicle in a Neural Network (ALVINN) [2]. The blockchain implementation proposed in this paper addresses the network structure and communications of participants and would not affect the Neural Network style operation of the AVs. Development in the AV field is very promising and is likely to alter our current and future transportation infrastructure. Ride-hailing and ride-sharing applications stand to benefit from utilizing new technologies as they would reduce both operating costs and the safety of the passengers in their network as AVs have been shown to operate at a higher safety standard than a human driver.

Blockchain was first introduced in 2008 when Satoshi Nakamoto [3] published which described a peer-to-peer electronic cash system known as Bitcoin. Since then, this concept of peer-to-peer cash systems has been developed in many variations and its underlying technology showed promising uses for the outside goal than original. Blockchain technology works as a distributed append-only ledger, where all information within the system is stored and accessed by connected peers. Lin *et al.* [4] outlines the different types of blockchain architecture, security issues, and challenges. Among various blockchain-based approaches, Hyperledger is a fully-permissioned network designed for operations involving sensitive and confidential data, whereas Ethereum is a public network [5].

**Contributions:** Our work intends to utilize Hyperledger Fab-ric to create a decentralized ride-hailing framework that would be beneficial to the development and adoption of a ride-hailing platform towards AVs because by nature blockchain technology creates trust between multiple non-trusting entities. "Drivers" are the owners of the AVs and would utilize this platform to form a network that would function as if created and maintained by a centralized source with all of the benefits of centralization. Trust between participants in the network is provided through the chaincode protocol that allows data security for differing client applications to participate in the network without fear of data theft or tampering. The implementation in this paper is transportation system agnostic in that it is not specific to AV ride-hailing and could be used as a decentralized ride-hailing platform for standard human driver ride-hailing.

The rest of this paper will be organized as follows: Section II provides background information including descriptions of related technologies and challenges that we overcome during the development of the framework described in this paper, Section III describes research that has already been done in related fields, Section IV describes the proposed framework, Section V provides detailed implementation of our framework, Section VI discusses a case study involving a simulation using real-world locations. Section VII provides a security evaluation of the framework, Section VIII demonstrates a load evaluation of our implementation, and Section IX concludes the paper and discusses future work.

## II. RESEARCH BACKGROUND AND MOTIVATION

This section will describe anticipated challenges with this work and our motivation for creating and implementing this framework. A brief background of blockchain technology and ride-sharing services is provided as well.

### A. Overview of Blockchain

A blockchain is essentially a chain of hashed 'blocks' where each block contains a time-stamp, the previous block's hash, and a collection of transactions. In other blockchain implementations, these transactions can be an invocation of code stored in the ledger known as 'smart contracts'. A block is generated after a set of transactions have been invoked and are awaiting validation.

There are different types of blockchains including (i) Public (ii) Private and (iii) Consortium Blockchain. In a public blockchain, miners participate in the consensus determination process and the ledger is completely visible to all participants. Public blockchains are permissionless and do not implement access control regarding transaction acceptance. While Private blockchains utilize a centralized architecture where one business or entity controls all of the nodes in the blockchain and writes and validates all transactions. This allows higher efficiency and strict permissions on who can participate in the network. However, all of the flaws that accompany centralization remain. On the other hand, consortium blockchain only allows trusted nodes to participate in the validation of blocks but these trusted nodes are not defined to a single organization or entity. This can provide some of the benefits of the private blockchain such as efficiency and privacy of transactions without compromising the decentralized nature of the public blockchain.

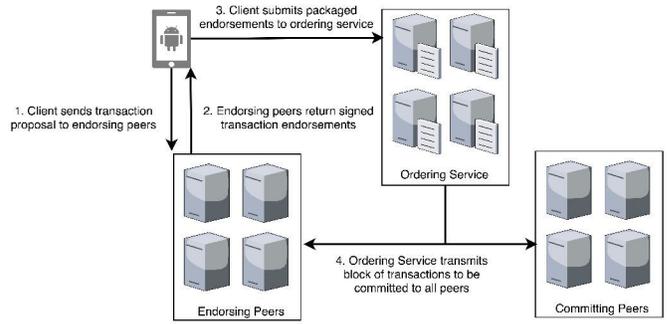

Fig. 1. Hyperledger Fabric Transaction Flow

### B. Hyperledger Fabric

A consortium blockchain technology known as Hyperledger Fabric [6] is utilized in this research. In Hyperledger Fabric, nodes must be certified before they can participate in the network. However, the nodes are not necessarily owned by one entity. Hyperledger Fabric supports smart contracts (termed "chaincode") that can be written in any programming language which defines all allowable interactions within the network. Each chaincode function has access to control functionalities such that only certain users / peers can invoke it.

When chaincode is installed on a peer it becomes an "endorsing peer" and endorsing peer validates a proposed transaction it returns an "endorsement" to the invoking user which contains the endorsing peers cryptographic signature to mitigate falsification. The user must receive a minimum number of endorsements (specified during chaincode deployment) prior to submission to the ordering service. The ordering service packages received transaction proposals into blocks according to a modular algorithm determined at channel creation. Blocks are sent to all peers participating in the channel to be committed to the ledger after one more check for validity. The complete flow from validating a transaction proposal to committing a block is illustrated in Fig. 1.

### C. Ride-Sharing / Ride-Hailing Background

The term ride-hailing describes companies such as Uber and Lyft where a rider requests a specific ride from their current location to a specified destination. The term ride-sharing formally defines situations where a rider accompanies a driver for a portion of a pre-planned trip that was being driven regardless. The implementation described in this paper is directed towards ride-hailing platforms but could be translated to a ride-sharing scenario.

Dynamic ride-sharing describes the problem space of routing for independent rides where routes must be calculated at the time of request rather than beforehand. This can be a challenging topic for optimization due to the lack of internal structure that other forms of ride-sharing such as buses and trains benefit from. There are many variables that must be taken into accounts such as ride distance, rider wait time, and the total number of rides given. The proposed framework does not directly contribute to the optimization of ride-hailing and routing algorithms rather underlying ride-hailing architecture.

*D. Research Challenges and Objectives*

Permissioned blockchain technology provides a solution by allowing AV owners to securely participate in the network while sharing the burden of maintaining the infrastructure. Hyperledger Fabric provides this functionality and our implementation intends to ease the adoption of this technology. In our framework, AV owners can join together to create Hyperledger Fabric organizations where they have control over factors such as infrastructure costs and profit distribution. The infrastructure becomes distributed in this manner and information security can be provided due to the dissemination of ride information being limited to peers participating in a transaction. Currently, transportation is mostly an independent consideration where owning a personal vehicle is a necessity outside of urban locations. Adoption of AVs would allow for the normalization of vehicle sharing which could, in turn, reduce the environmental impact of the automotive industry, provide a financially practical form of transportation to users, and generate profits for AV owners.

Information sharing within the network between actors must be based on the necessity for the sake of security. Riders should not have the authorization to access information related to rides that they did not actively participate in. In a co-rider scenario, each rider should only have access to information that they were present to observe. For example, co-rider pickup and dropoff locations should only be accessible to a rider if they were present when the information was recorded. Maintaining a ride-hailing platform independently is costly and with the addition of AVs, it becomes prohibitively expensive for most. The alternative architecture proposed in this paper allows individuals to lease their AVs to a community pool where service can be provided to a larger network of users.

## III. RELATED WORK

This section will be used to detail the research that has previously been done in this field or closely related fields.

*A. Ride-Sharing Algorithm Optimization*

The algorithmic matching of drivers to riders is an active research topic as the optimal matching algorithm has yet to be developed. Li *et al.* in [7] present an enhanced ride-sharing matching algorithm that takes into account meeting points and preferable time windows. Geisberger *et al.* in [8] approach this problem from the perspective of optimizing ride selection by selecting detours that maximize the serviceable area and minimize non-optimal driving routes. Lin *et al.* in [9] propose a ride-matching algorithm that optimizes based on factors such as travel mileage and time wasted waiting or riding unnecessarily which achieves a 16 percent mileage reduction and 66 percent driver availability increase in simulation. Xing *et al.* in [10] propose a ride-sharing system optimized for short rides in highly urban areas. Optimization of the ride-sharing problem cannot often be done in a generalized fashion due to the many variables within the environment.

There are many publications regarding the optimization of dynamic ride-sharing (DRS) algorithms. Agatz *et al.* in [11] formally define DRS and the challenges faced when developing optimized algorithms in a DRS system. Agatz *et al.* in [12] go on to present a dynamic ride-sharing approach optimized for minimum system-wide vehicle miles and transportation costs. The lack of rides is a common challenge in DRS systems which optimize for minimum vehicle miles. Shen *et al.* in [13] describe a dynamic ride-sharing framework that expands upon previous real-time ride-sharing algorithms. Kleiner *et al.* in [14] propose a mechanism for increasing the probability of users finding rides in urbanized environments with a DRS system while maintaining low system-wide vehicle miles.

Another relevant research area that is being explored is multi-hop ride-sharing systems that allow riders to transfer between drivers within a single ride which is more flexible in the scheduling of rides and for the reduction in the amount of total detours in the system. Drews and Luxen in [15] propose an algorithm for performing multi-hop routing in a ride-sharing system by utilizing time table graph search similar to that seen in urban metro systems. Teubner and Flath in [16] discuss the competitiveness of multi-hop ride-sharing systems when compared to more traditional and simplistic ride-sharing systems.

*B. Ride Hailing Privacy*

Traditional ride-hailing services such as taxis did not require strong privacy protection as riders could remain relatively anonymous during transactions. With the advent of application-based ride-hailing, privacy is a much larger concern, Pham *et al.* in [17] propose a framework for preserving the location privacy of riders and drivers without compromising on functionality. The framework was expanded [18] by increasing privacy and addressing the issue of user accountability that can be abused with anonymity. AÃ¯vodji *et al.* in [19] propose a privacy-preserving ride-sharing system that protects the privacy of users from the service provider during the matching phase of the ride-sharing system. Implementation of the privacy-protecting protocols proposed in these frameworks will be important to the framework proposed in our paper due to the decentralization of the service provider.

*C. Smart Contract Security*

MJH Faruk *et al.* in [20] propose a blockchain-based system for storing and sharing electronic healthcare records (EHR). The proposed application was implemented within the Hyperledger Fabric because of its permissioned nature and access control are implemented through chaincode permissions. Delmolino *et al.* in [21] discuss common smart contract development pitfalls as well as their smart contract security education efforts. This author also discusses some smart contract programming pitfalls that are common to any smart contract development such as logical errors and a lack of data encryption. Ethereum specific mistakes are presented in this paper as well as techniques to avoid/correct them. Luu *et al.* in [22] discuss common avoidable vulnerabilities in Ethereum such as Transaction Ordering Dependence, Timestamp Dependence, Mishandled Exceptions, and the Reentrancy Vulnerability. While some of these vulnerabilities are not directly tied to Hyperledger Fabric it can be useful to learn how mistakes are exploited in other blockchain environments to learn how to better protect a permissioned blockchain.

### D. Blockchain Ride-Sharing

In [23], Mehedi Hasan *et al.* describe a framework for an multi-purpose dependable blockchain to be used as the communication platform for an autonomous vehicle ride-sharing system. There are several cryptocurrency-based decentralized ride-sharing efforts either currently in development or that have been developed and are in the market as of right now, such as [24], [25], [26]. These projects are similar in design to the project described in this paper with the main difference being that these projects are all public blockchain implementations, mostly Ethereum-based. Public blockchains are not ideal for this work due to the need for private information to be shared between smart contracts. This paper uses Hyperledger Fabric as its blockchain

## IV. FRAMEWORK

### A. Architecture

The core components of the Hyperledger Fabric architecture that support the ride-hailing framework proposed in this work are: Organizations, Endorsing Peers, Channel Structure, Chaincode Function Structure, Driver / Rider Clients, and Certificate Authorities. The Hyperledger Fabric organizations used in this framework are created and maintained by groups of drivers / AV owners. These organizations have several core components and this section will detail those components and how they interact to provide ride-hailing functionality while protecting the confidentiality of users. Fig. 2(a) illustrates each of the components required by an organization in the proposed framework.

*1) Peer Nodes:* Organizational peer nodes will act as both endorsing peers and committing peers. Endorsing peers are sent transaction proposals from driver and rider client applications in the network and return signed proposal responses. Proposal responses are signed cryptographically to minimize falsification and are marked accepted or rejected based on the transaction validity. After transactions have been ordered and validated, all committing peers in the channel commit the transaction to their local ledger.

*2) Certificate Authority / Orderer Node:* The root certificate authority of an organization issues certificates to all of the entities within the organization which are the orderer node(s), the peer nodes, and all of the end-user client applications (drivers and riders). These certificates are stored locally on each individual entity within the local MSP. Upon creation of the ordering service, all ordering nodes share local MSP details between one another. This allows all of the organizations in the network access to the certificates that can be used to validate the identity of entities in the network. Any certificate authority implementation can be used for this purpose but a client / server architecture is recommended by Hyperledger Fabric.

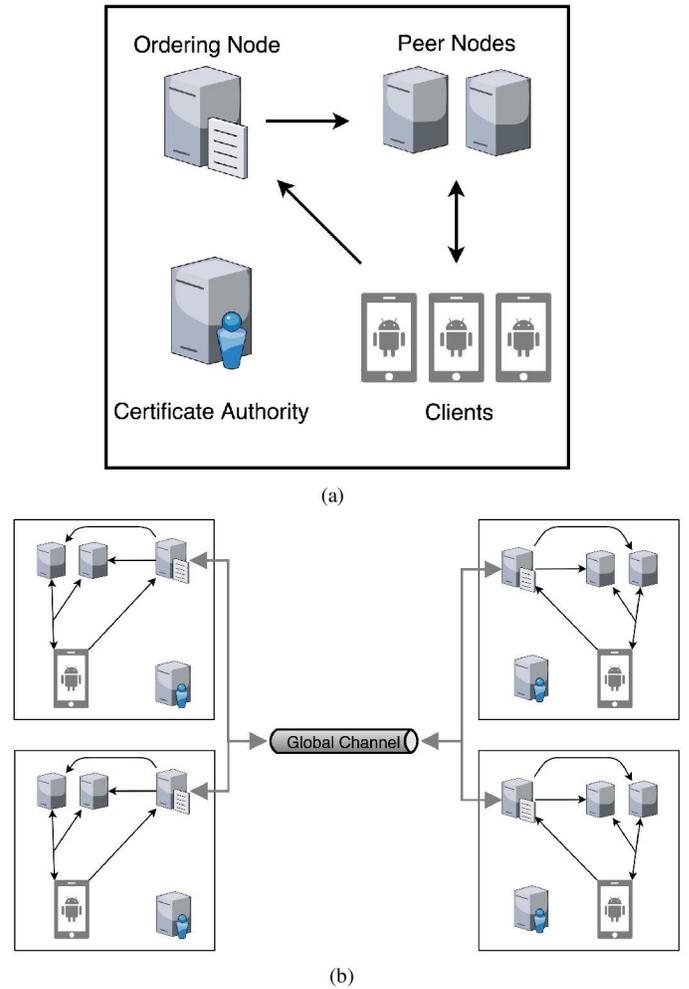

Fig. 2. Proposed Hyperledger Fabric Implementation: (a) Components of a Hyperledger Fabric Organization and (b) Communication Structure in Multi-Organization Hyperledger Fabric Network

### B. Chaincode Protocol

Careful design of the chaincode functions that are installed on endorsing peers is critical to the goal of securely storing rider information in a manner where it is accessible only to the rider it pertains. The distinction between individual riders and drivers is done within the chaincode by creating unique UserIDs and then utilizing these IDs within the chaincode functions. The specific implementation used for this work will be described in the implementation section below and an illustration of the protocol is illustrated in Fig. 3. The following functions provide the core functionality of the proposed framework and will be called by the client application by both drivers and riders to facilitate the ride-hailing process:

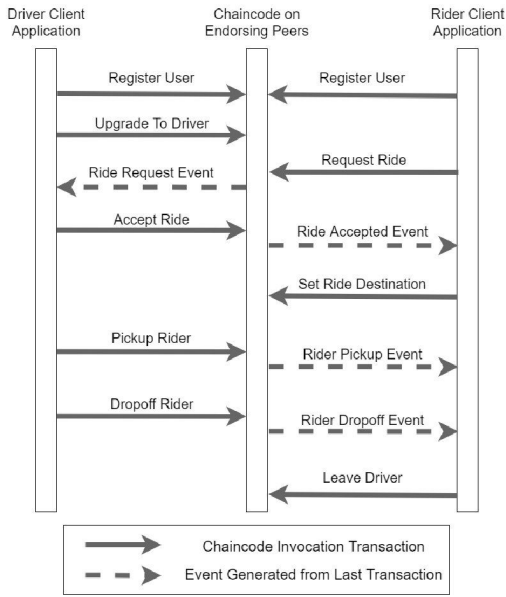

Fig. 3. Ride-Hailing Protocol Framework in UML Diagram

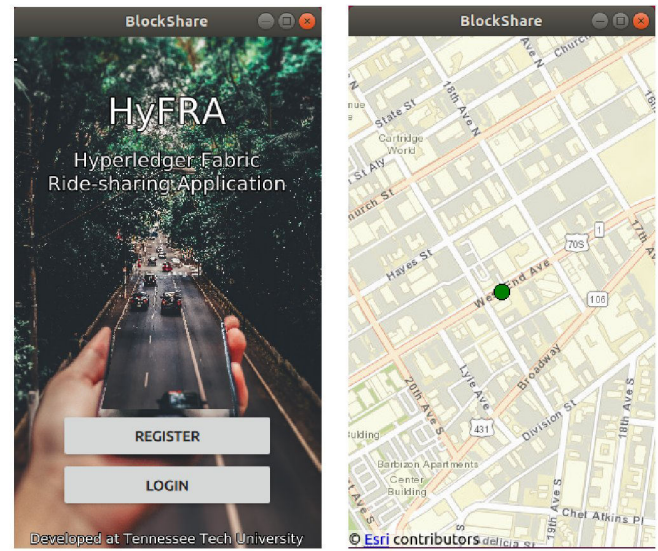

Fig. 4. Proposed Application Screenshots (HyFRA)

*1) RegisterUser / UnregisterUser:* Creates a new user object in the ledger using the unique UserID as the key and the function parameters for values. The values that are attached to a unique user are the hash and salt of the rider's password and an array of ride structs which will be filled as the user provides or requests rides. Unregister users need to delete this key from the ledger. A UserID will be associated with the driver's local MSP ID with the organization's global MSP ID that will ensure the security of user information assuming the chaincode framework is properly used and login functionality is implemented in the client application.

*2) RequestRide:* Creates a temporary ledger value for the ride that is being requested where data will be stored until the end of the ride. Each participant will retrieve the information relevant to them to be stored permanently in the ledger at the end of the ride. This function also needs to create an event that will be received by listening drivers (implemented in the client application).

*3) AcceptRide:* Updates the temporary ride object created by the previous function to mark that the ride has been accepted and to designate the accepting driver. The function creates an event which will alert the requesting rider that a driver is en route.

*4) SetRideDestination:* Updates the temporal ride object to include the ride destination coordinates when the signal from acceptRide is received. This is done after driver acceptance to prevent discrimination based on dropoff location.

*5) PickupRider:* Called when the driver reaches the rider's location. This function performs checks to ensure the ride is still ongoing and the driver is at the correct location and then triggers an event to alert the rider that the driver has arrived.

*6) DropoffRider:* Called when the driver reaches the final destination. Pulls necessary information from the temporal ride object to create a permanent ride object specific to the driver and creates an event to alert the rider that the ride is ending.

*7) Other Functions:* The proposed framework will also consist of various functions; for instance, leaveDriver, setCoriderInformation, and getUserInfo will create a permanent ride object for the rider, to check if the co-rider joins or leaves the ride, and to retrieve the user's password hash, list of RideIDs respectively. Other important functionality would need to be added to align with the usability of the users.

## V. IMPLEMENTATION

Our proposed framework is implemented using two organizations both having two peers, an orderer, a certificate authority, and then multiple user accounts which are utilized by the client application to register drivers and riders. It was decided that organizations should maintain a minimum of two peer nodes so that a singular organization can operate independently while still providing fault tolerance. There is not a formal certificate authority node within the organizations in this implementation but rather cryptogen (the certificate tool provided by Hyperledger Fabric) was integrated into the build process. The client application is built using Golang and we utilize the fabric-sdk-go library [27] to communicate with the endorsing peers of our network.

The peers of this network are built using docker compose [28] and run within separate docker [29] containers in a single docker network. The client application currently runs on the host machine and accesses the Hyperledger Fabric network through ports that have been exposed to the host machine via docker. In a production deployment the peer, orderer, and CA nodes would be servers accessible via WAN. These servers can all be located within the same machine and could still run in docker (which is ideal for fault tolerance) but they need to be accessible via defined sockets separate from one another. The client application GUI was built for the Qt cross-platform application development library [30] [31] so that application could be developed for desktop and easily ported to android devices.

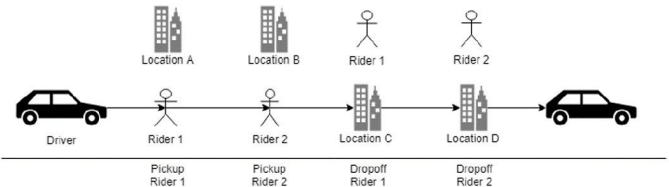

Fig. 5. Simple Co-Rider Situation

TABLE I
CASE STUDY LOCATION COORDINATES

| Location | Latitude | Longitude |
|---|---|---|
| Nashville International Airport | 36.13149 | -86.6694 |
| Nissan Stadium | 36.16624 | -86.7719 |
| Nashville Greyhound Station | 36.15212 | -86.7735 |
| Belmont University | 36.13515 | -86.7955 |

After running the client application, the user is presented with the main menu with basic login functionality as shown in Fig. 4(a). A user must be registered before any additional functionality can be utilized. The login functionality of the client application prompts the user for their username, organization, and password. The rider is also given an option to authenticate as a rider or a driver. When a user registers they are automatically created as a rider in the ledger. An option to upgrade to a driver is given after login. If the user is logging in as a driver they are presented with a startDriving function and a logout function which will return the user to the main menu where they will need to log in again to access the ride-hailing functionality. The startDriving function is the core of the driver's client application experience and starts the ride-hailing process. Before ride requests can be successful at least one driver must be driving. After the startDriving function is activated the user is prompted to provide the address of their current location. The geolocation service and mapping service are used for latitude and longitude. We utilized a plugin for Qt which interacts with the open-source project named Open Street Map [32]. Once a latitude and longitude for the driver have been obtained, the driver's application updates with a map of their current location and the driver start listening for events on the requestRide chaincode. The starting driver map can be seen in Fig. 4(b).

When a rider authenticates they are presented with the option to request a ride, update their profile, upgrade to driver status, or log out. After listening to the events on the acceptRide chaincode function, the rider sends their destination coordinates to the setRideDestination chaincode function where the temporal ride object will be updated. Upon arrival, the driver's client application updates the temporal ride object. As the driver is moving from the rider's pickup location to the newly added dropoff location the driver listens for new ride requests. When a ride ends the dropoffRider chaincode function is called which finalizes ride information and generates an event with the RideID. All co-riders use the RideID to update their co-rider dropoff location in their temporal ride object.

## VI. CASE STUDY

Here we discuss a synthetic case study demonstrating the execution of our blockchain-based ride-hailing platform.

### A. Simple Co-rider Scenario

A simple ride-hailing situation is depicted in Fig. 5 where two riders ($R_1$ and $R_2$) request a ride from driver $D_1$. In this scenario $R_1$ is present for $R_2$'s pickup and $R_2$ is present for $R_1$'s dropoff. This information should be reflected during the execution of transactions within the Hyperledger Fabric network. The resulting ledger query from $R_1$ should show the pickup location for $R_2$ but not the dropoff location and vice versa for the query by $R_2$.

The locations used for this scenario are all public locations in Nashville, TN, and represent a typical use case for the ride-hailing platform's operation. The driver starts listening for new ride events while located in downtown Nashville. $R_1$'s starting location is Nashville International Airport and his destination is Nissan Stadium. $R_2$'s starting location is the Nashville Greyhound station and their destination is Belmont University. The latitudes and longitudes for these locations are illustrated in Table I. The locations are stored in latitude/longitude pairs in the ledger so this table will be important as a reference.

When the driver elects to start listening for ride requests they are shown a screen similar to Fig. 4(b) with a marker on their current location. In a final build of the application, this screen would update with the location pulled from the mobile device GPS. For development purposes, the coordinates are entered when driving is initiated and remain static until a ride is received. At this point in time, the driver has authenticated with the Hyperledger Fabric network and registered for chaincode events on the requestRide chaincode function.

The next step for this scenario is $R_1$ opts to request a ride in their application. The rider enters a starting address and ending address which is converted to latitude and longitude using geolocation. The rider takes this latitude and longitude and calls the requestRide chaincode where the temporal ride request key in the ledger is created which can be seen in Table II. As seen in Table II the ID of the key in the ledger is built using values specific to the rider. Org2PeerOrgMSP is the unique MSP name and eDUwOT is the rider's unique ID within the MSP. For this specific ride request to be generated this user must call the chaincode because these two values are generated using the certificate that is passed by the calling user which s the key part of how the security of this architecture functions. When a driver accepts the ride the status field of the temporal ride request object is updated as "accepted" which allows for other drivers to be notified.

In this scenario when the driver arrives at $R_1$'s pickup location and begins moving towards the destination another ride is requested by $R_2$. The driver prompted his application to either accept or deny this ride. In this case study, the driver accepts the ride which updates the current destination. After reaching $R_2$ for pickup, the location must be stored for $R_1$ so that it can be referred to later in the permanent copy of this ride's key. $R_1$ was present for $R_2$'s pickup so this information should be accessible to $R_1$.

TABLE II
TEMPORAL RIDE REQUEST CREATED IN LEDGER WHEN RIDER-1
REQUESTS A RIDE AND COMPLETED TEMPORAL RIDE REQUEST FOR
RIDER-1 BEFORE DELETION

| RideRequest | Org2PeerOrgMSP (ID) | eDUwOT |
|---|---|---|
| RideID | ID-eDUwOT | ID-eDUwOT |
| DriverID | N/A | ID-06Q049V |
| Status | Requested | Completed |
| PickupLocation | 36.1452/-85.4969 | 36.1452/-85.4969 |
| DropoffLocation | N/A | 36.17488/-85.5089 |
| PicupTime | N/A | 12/5/2018 12:34 |
| DroopoffTime | N/A | 12/5/2018 12:36 |
| Co-RiderID | N/A | ID-XNIcjF |
| Co-RiderPicLocation | N/A | 36.15395/-85.5138 |
| Co-RiderDropLocation | N/A | N/A |

Table II shows $R_1$'s temporal rideRequest object prior to relocation to permanent storage and then deletion. The Co-rider Pickup Location field has been filled because $R_1$ was present for this portion of the ride and therefore the information is relevant to $R_1$. $R_2$'s final temporal rideRequest was very similar with the exception that $R_1$'s dropoff location was recorded. The driver handles the archiving of co-rider pickup and dropoff locations because he has access to all of this information during the ride. After the ride is over the riders can only access information that is directly relevant to them. This architecture can be carried over to other applications as well or could be expanded within this application to store additional sensitive data.

## VII. SECURITY EVALUATION

We evaluate the proposed framework and implementation to determine its effectiveness, resiliency, and preservation of sensitive information. An adversarial model is used to define what possible vulnerabilities and a formal audit of the code using a static analysis tool is used to show how these vulnerabilities have been mitigated.

### A. Adversarial Model

The following three adversaries are defined for the analysis of the ride-hailing system:
*User, Operator, and Outsider:* User refers to an active adversary who might attempt to manipulate transactions to either retrieve sensitive information pertaining to another User of the system or receive monetary benefits in the form of cheaper rides, disproportionate payment, or theft. While Operators control the network hardware who might try to manipulate the network structure to receive a disproportionate portion of requested rides. Besides, an Outsider is a passive adversary who might attempt to sniff network traffic to identify sensitive information to blackmail, impersonate, steal from, or otherwise harm a Users.

### B. Threat Taxonomy and Proposed Mitigation

*1) Sybil Attack and Eclipse Attack:* We take consideration to detailing with different attack vectors and the respective mitigating factors in our implementation including Sybil Attack and Eclipse Attack. One concern in decentralized networks is an eclipse attack where an attacker isolates an entity to where the victim cannot participate in the network at large. In the proposed architecture this would mean all transactions sent by a client would be endorsed only by a subset of peer nodes. This can be avoided by utilizing an endorsement policy that requires an endorsement from at least one peer node from a different organization.

*2) Malicious Ledger Query and Packet Sniffing:* We also evaluate Malicious Ledger Query where users may attempt to retrieve various information about another entity by querying the ledger. This attempt is mitigated in our framework by the unique chaincode design where ledger queries are not allowed unless invoked through a chaincode transaction. An outsider might try to sniff the networks that transactions travel through to retrieve or infer private information about a user which call Packet Sniffing. TLS [33] can be configured to work with all network communication between entities in the Hyperledger Fabric network. TLS is an industry-standard encryption technology used for information in transit and can be trusted for this purpose.

*3) Malicious Client Application:* An Operator may attempt to develop a client application that could be used to harvest user information such as passwords or credit card information. Unfortunately giving the control of client application development to the organizations creates this possibility; however, without a client's certificate as well this information could not be used. Also, a malicious client application could not interact with the network because of chaincode version requirements.

### C. Chaincode Analysis

We utilized a static analysis tool for Hyperledger Fabric chaincode called ChainSecurity [34]. We used design patterns in non-determinism or states that could be exploited by malicious actors in the system. The chaincode deployed in our implementation did not have any flaws according to the analysis done by this tool. This static analysis tool checks for concurrency, unchecked exceptions, ledger operations that depend on the global state, field declarations, blacklisted import statements, reading from the ledger after a write operation, and unsanitized input to the chaincode. The global state and field declarations being used in ledger operations could be exploited because these global variables are only specific to the peer the chaincode is installed on. If this peer crashes or becomes out of sync with the other peers of the network it will never be able to submit another transaction due to its improper readset. The static analysis of the chaincode assures that the implementation will not be susceptible to these types of attacks.

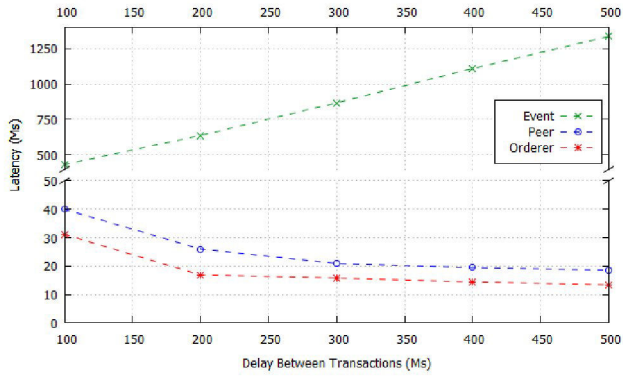 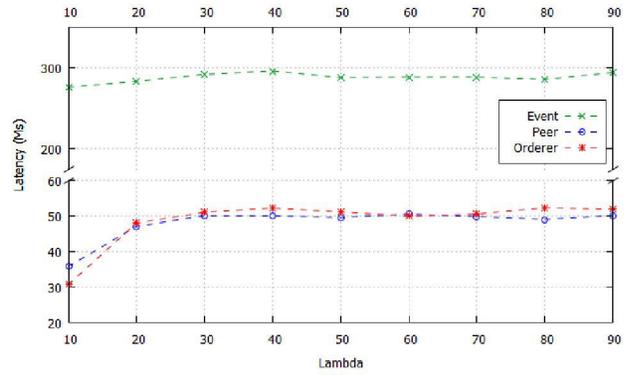

Fig. 6. (a) Constant Rate Traffic Load Test and (b) Poisson-Based Traffic Load Test

## VIII. Load Resiliency Analysis

We utilized a tool provided by Hyperledger Fabric within the fabric-test repository [35] called the Performance Traffic Engine (PTE) to test our implementation under load. PTE is designed to enable testing live Hyperledger Fabric networks with various chaincode, orderers, and peers. All test cases processed a total of 6,000 transactions for a total of 1,000 rides with varying transaction rates. Transactions were sent at varying rates according to the test distribution but were handled by four processors with two processors dedicated to each organization. Each data point in Fig. 6(a) and Fig. 6(b) reflects a single test case. All transactions and events sent were received successfully across all tests. The ordering service used in our implementation has a batch timeout of 2 seconds and a max message count of 10. Whenever one of these limits is reached a block is created.

There are three values being recorded across all tests: peer, orderer, and event latency. Peer latency measures the amount of time between the moment a transaction proposal is submitted for endorsement and the moment the endorsements are returned to the client. The orderer latency measures the amount of time it takes to receive a transaction acknowledgment from the ordering service after the client submits its endorsements. Event latency measures the time between event registration before submission to the ordering service and a successful block commit. Each value is averaged over 1000 transactions and these averages can be seen in the graphs below.

### A. Constant Rate Network Traffic

The first set of tests sent transactions to the network at varying constant rates with a 30% deviation. Fig. 6(a) shows the results of sending traffic to the network with the minimum delay being 100ms between transactions and with a maximum delay of 500ms. As the delay between transactions increased the event latency also grew at a constant rate. The event latency is measured end-to-end between transactions so the increase is logical due to the artificial delay between transactions. Orderer and peer latency showed an overall decrease as the delay between transactions increased due to the decreasing load being placed on the network.

A second constant rate load test was done with delays ranging between 10 and 90 ms. The event, peer, and orderer latency all remained essentially constant throughout the test due to transactions being sent from one machine using 4 processors. The delay between transactions was not large enough for previous transactions to be received by the testing engine. As the tests approached a delay of 90 ms the peer and orderer delay can be seen to slightly decrease as a more reasonable testing speed is approached.

### B. Poisson Distribution-Based Network Traffic

In the interest of testing the network under different types of loads, the next test sent transactions according to the Poisson distribution with varying lambda values. The Poisson distribution was chosen because it relates to events that occur independently and is often used to model event functions where the average number of events is known. In our scenario, we tested scenarios with varying transactions per second as the average event variable lambda.

Fig. 6(b) shows the results of testing the network with a lower range of lambda values between 10ms and 90ms. Event latency remained mostly constant as the lambda value increased due to the network load increasing and the testing delay decreasing which had the effect of balancing one another. Event and peer latency increased as transaction rate increased due to the increased load on the network which was the expected result. After reaching a lambda of 30 the peer and event latency leveled off and remained constant. A higher lambda test was conducted as well with lambda ranging between 100 and 1000 but all of the measured latencies remained constant again due to the limitations of system resources.

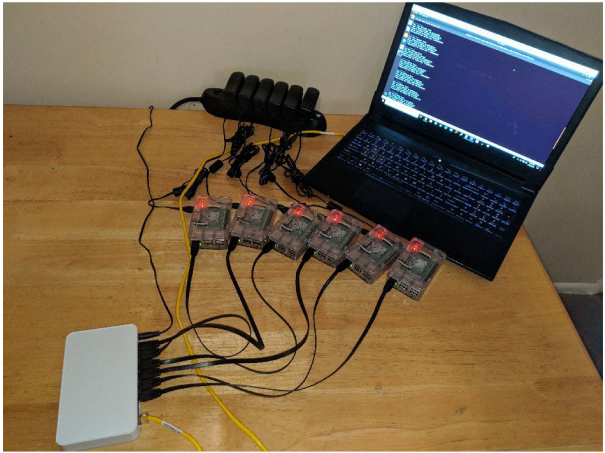

Fig. 7. Raspberry Pi Test Bed

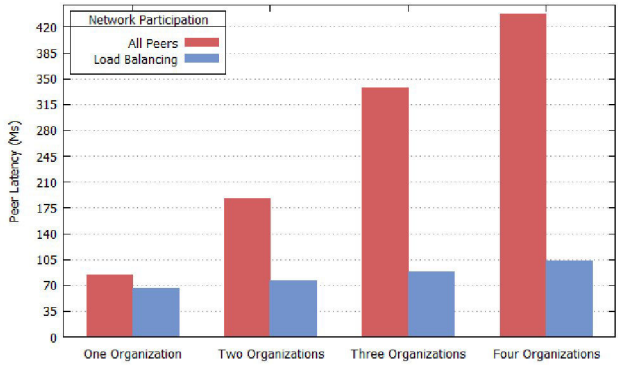

Fig. 8. Peer Latency with Differing Numbers of Peers

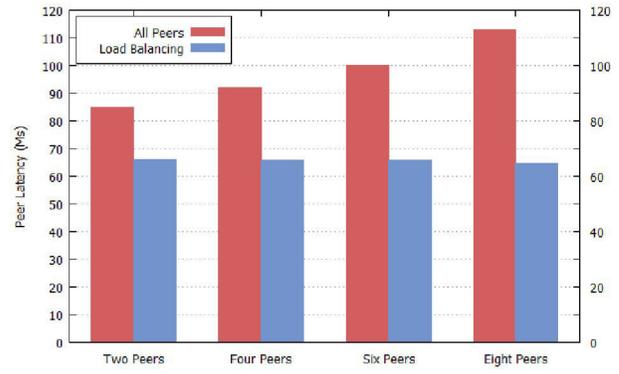

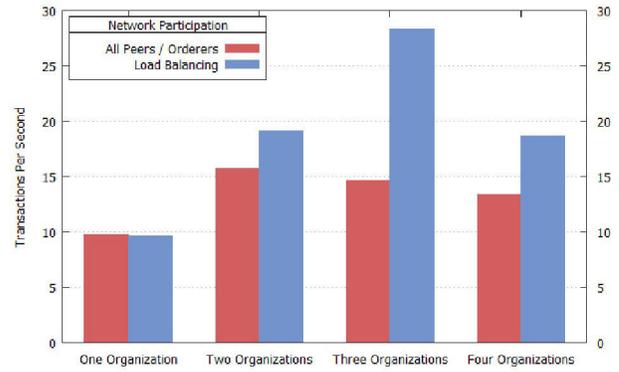

Fig. 9. (Network Statistics of Organization Reconfiguration Tests: (a) Peer Latency and (b) Transactions Per Second.

## C. Organization Restructuring Tests

The final set of tests measured the effect of having different configurations of peer and orderer nodes as well as the effect of having more organizations participate in a network. These tests required more computers so that the results were not limited by the processing power of a single machine. To complete these tests a test-bed was built using Raspberry Pi computers [36] as the host machines for the organizations in the testing. A picture of the test-bed is shown in Fig. 7. Each Raspberry Pi was running a Ubuntu 18.04 image that was designed for ARM architectures. In each of the tests, one of the Raspberry Pi's acted as a dedicated Kafka / Zookeeper node cluster which the ordering nodes of the network communicated. The Raspberry Pi machines were connected via a network switch to the testing computer where transactions originated from and sent in bursts with a delay of 200ms between them.

The first test measured the effect of joining additional peers to a single organization while the amount of traffic being sent remained constant. Neither the speed nor amount of traffic sent to the organization was adjusted between tests and the results can be seen in Fig. 9. Two separate tests were conducted, one required all peers to endorse each transaction, and the other balanced endorsements between peer nodes. In the first scenario, the peer latency was increasing over time because of the communication overhead that is required by the client to organize the endorsement of their transactions through each peer. The second test utilized peer load balancing where only one endorsement from a peer node was required for a transaction to be considered valid. The peer latency slightly reduces over time because there are more peers than there are transactions being received and therefore the workload on a single peer is never increasing.

The final test increased the number of organizations as well as the amount of traffic in the network at each step to measure the effect on performance. The simulation was again split into two overarching tests where one required all peer nodes to participate in the validation of each transaction and one where the load was distributed evenly between peer nodes. The peer latency measurements can be seen in Fig. 8 where it increases but only by 10-15 Ms at each step. Fig. 9(b) illustrates the transactions per second (TPS) that were being committed to the ledger. It can be seen again that the load balancing simulation outperforms the full participation simulation. In both tests, there is an increase in TPS before it begins to fall. The increase can be attributed to the network not being fully utilized until the network traffic rate is increased to the optimal level. The decrease in TPS comes after the network reaches a critical mass of transactions after which performance drops off. The point was reached in both tests where the peer nodes could no longer process transactions before the next set of transactions were received which formed a bottleneck during the endorsement stage. These transactions are being endorsed on the same machines that must commit the blocks after the ordering stage. Work is constantly being done on the peer node which slows down the performance of the committing stage where additional verification is done. The only way to subvert this bottleneck would be to segregate the endorsing peers from the committing peers which would easily be possible with this architecture; however, it would increase hardware costs for the organization operators.

## IX. Conclusion

This paper serves as a framework for building a decentralized ride-hailing application that serves as the intermediary platform connecting drivers and riders. The chaincode protocol described in this paper provides security of transactions through design and could be extended to many applications. The implementation utilizes permissioned nature and built-in organization structure of Hyperledger Fabric to detail an optimal build for organizations of independent drivers. Ideally, each organization in this system will have the ability to create their own client application and still be able to interact with the Hyperledger Fabric network and share the load of the riders. Certificate authority public key infrastructure is recommended by Hyperledger Fabric but future development could utilize a peer-to-peer public key infrastructure that works using a "web of trust" to add another layer of transparency and decentralization to the model proposed in this paper.

## Acknowledgment

The work was primarily supported by the U.S. National Science Foundation (NSF) award #1565562, and partially by NSF awards #1723578 and #2100115. Any opinions, findings, and conclusions or recommendations expressed in this material are those of the authors only.